\documentclass[10pt]{iopart}

\usepackage{upgreek}
\usepackage{graphicx}
\usepackage{subcaption}

\usepackage[colorlinks,
            citecolor=red,
            urlcolor=blue,
            bookmarks=false,
            hypertexnames=true]{hyperref}

\usepackage[normalem]{ulem}

\usepackage{color}
\definecolor{brickred}{rgb}{0.8, 0.25, 0.33} 

\begin{document}

\title[]{High kinetic inductance NbTiN films for quantum limited travelling wave parametric amplifiers}.

\author{F.~Mantegazzini$^{*,1,2}$,
        F.~Ahrens$^{1,2}$,
        M.~Borghesi$^{3,4,5}$,
        P.~Falferi$^{1,6,2}$,
        L.~Fasolo$^{7}$,
        M.~Faverzani$^{3,4,5}$,
        E.~Ferri$^{4}$,
        D.~Labranca$^{3,4,5}$,
        B.~Margesin$^{1,2}$,
        R.~Mezzena$^{8,2}$,
        R.~Moretti$^{3,4,5}$,
        A.~Nucciotti$^{3,4,5}$,
        L.~Origo$^{3,4,5}$,
        A.~Vinante$^{1,6,2}$,
        M.~Zannoni$^{3,4,5}$,
        A.~Giachero$^{*,3,4,5}$
}
\address{$^{*}$ \textit{Corresponding author}}
\address{$^{1}$ Fondazione Bruno Kessler (FBK), I-38123 Trento, Italy}
\address{$^{2}$ INFN - TIFPA, I-38123 Trento, Italy}
\address{$^{3}$ Dipartimento di Fisica, Universit\`{a} di Milano-Bicocca, I-20126 Milano, Italy}
\address{$^{4}$ INFN - Sezione di Milano Bicocca, I-20126 Milano, Italy}
\address{$^{5}$ Bicocca Quantum Technologies (BiQuTe) Centre, I-20126 Milano, Italy}
\address{$^{6}$ IFN-CNR, I-38123 Trento, Italy}
\address{$^{7}$ INRiM - Istituto Nazionale di Ricerca Metrologica, I-10135 Turin, Italy}
\address{$^{8}$ University of Trento, I-38123 Trento, Italy}

\ead{fmantegazzini@fbk.eu}
\ead{andrea.giachero@mib.infn.it}

\vspace{10pt}
\begin{indented}
\item[]
\end{indented}

\begin{abstract}
A wide-bandwidth and low-noise amplification chain in the microwave regime is crucial for the efficient read-out of quantum systems based on superconducting detectors, such as Microwave Kinetic Inductance Detectors (MKIDs), Transition Edge Sensors (TESs), Magnetic Microcalorimeters (MMCs), and RF cavities, as well as qubits. Kinetic Inductance Travelling Wave Parametric Amplifiers (KI-TWPAs) operated in a three-wave mixing fashion have demonstrated exceptional dynamic range and low-noise performance, approaching the quantum limit. These amplifiers can be fabricated using a single layer of a high kinetic inductance film as weakly dispersive artificial transmission lines, with the ability to control the phase-matched bandwidth through dispersion engineering. In this study, we present the optimisation of the rf sputter-deposition process of NbTiN films using a Nb$_{80\%}$Ti$_{20\%}$ target, with the goal of achieving precise control over film characteristics, resulting in high kinetic inductance while maintaining a high transition temperature. The parameter landscape related to the different sputtering conditions, such as pressure, power, and nitrogen flow, has been explored and the film thickness has been used as a fine-tuning parameter to adjust the properties of the final NbTiN films used for the fabrication of KI-TWPAs. 
As a final result, we have obtained a NbTiN film with a kinetic inductance of 8.5\,pH/sq which we have exploited to fabricate KI-TWPA prototype devices, showing promising amplification performance.
\end{abstract}

%
%
\submitto{Physica Scripta}

\section{Introduction}

Ultra low-noise microwave detection plays a central role in many advanced applications. In particular, a broad-band and low-noise amplification chain is a key element for quantum systems and quantum technologies, e.g.~multiplexed read-out of qubits~~\cite{Bronn2017}, quantum key distribution~~\cite{Fesquet2022} and microwave quantum illumination~~\cite{Fasolo20221}, and for fundamental physics experiments, e.g.~axionic dark matter search~~\cite{Braggio2022, Peng2022}, dark photons detection~~\cite{Ramanathan2022}, and microwave multiplexed read-out of particle and astro-particle detector arrays, such as microwave kinetic inductance detector (MKIDs)~~\cite{Zobrist2019}, transition edge sensors (TESs)~~\cite{Alpert2019} and magnetic microcalorimeters (MMCs)~~\cite{Kempf2013}. 

The first amplification stage sets the noise performance of the amplification chain~~\cite{Friis1946} and currently, in cryogenic measurement set-ups, it is typically implemented with state-of-the-art commercial high electron mobility transistor (HEMT) amplifiers. These amplifiers provide a high gain and a large bandwidth, but with added noise 10–25 times above the standard quantum limit~~\cite{Caves1982,Clerk2010}, the fundamental limit imposed by quantum mechanics. 
Instead, superconducting parametric amplifiers offer the possibility to exploit highly non-linear superconducting elements without introducing dissipation into the system, thus suppressing thermal fluctuations and eventually approaching the quantum limit set by zero-point quantum fluctuations~~\cite{Clerk2010}. 

Resonant Josephson parametric amplifiers (JPAs), based on the fast modulation of the dissipationless and non-linear inductance of a Superconducting Quantum Interference Device (SQUID), have already demonstrated parametric gain with quantum limited noise ~\cite{Castellanos2007, Aumentado2020}. However, despite efforts to increase the bandwidth up to a few hundred megahertz via impedance engineering~~\cite{Mutus2014,Roy2015}, JPAs are ultimately limited by the constraints set by the resonant cavity that limits the achievable bandwidth, making broadband amplification essentially impossible to achieve. Recently, impedance matched Josephson parametric amplifiers (IMPA) based on the rf-SQUID, have achieved a bandwidth of 250–300$\,$MHz~~\cite{White2023}. Despite this clear improvement, the developed amplifiers still cannot read out more than six devices simultaneously.
A narrow bandwidth particularly hinders applications such as microwave multiplexed read-out of high number of cryogenic detectors or qubits, or broadband squeezing generation.

In order to overcome these limitations, travelling-wave Josephson parametric amplifiers (JTWPAs) based on transmission lines with embedded Josephson junctions have been successfully developed, achieving near quantum-limited noise performance ~\cite{Macklin2015}.
However, both for JPAs and JTWPAs, the presence of Josephson junctions with critical currents in the order of few microampere inevitably constraints the dynamic range of such devices, limiting their usage for the applications mentioned above. The dynamic range can be measured considering the 1dB compression point, which increases proportionally to the dissipation rate set by the coupling to the transmission line. Typical values of the 1dB compression point for optimised Josephson junction-based parametric amplifiers are in the order of $-100 \, \mathrm{dBm}$ for $20 \, \mathrm{dB}$ gain \cite{Eichler2014, Frattini2018}.
Moreover, JTWPAs involve multi-layer and complex fabrication processes which hinders their wide adoptions in laboratories.  

These considerations clearly motivate the development of a new type of broadband travelling-wave parametric amplifiers, called Dispersion-engineered Travelling Wave Kinetic Inductance (DTWKI) parametric amplifiers or Kinetic Inductance Travelling Wave Parametric amplifier (KI-TWPA or in short-form KIT), based on the nonlinear kinetic inductance of superconducting transmission lines, as recently proposed~\cite{Eom2012}. 
The DARTWARS (Detector Array Readout with Traveling Wave AmplifieRS) project~\cite{DARTWARSweb} aims to develop large bandwidth travelling-wave parametric amplifiers that operate at the quantum limit and provide high gain, high saturation power, and a large dynamic range, exploiting both technologies based on Josephson junctions~\cite{Rettaroli2022} and high kinetic inductance films~\cite{Borghesi2022}. 

KI-TWPA devices leverage the intrinsic non-linearity of the superconducting film and, thus, the fine-tuning of the film characteristics is crucial. In particular, the critical temperature, the sheet resistance and the kinetic inductance of the film must be under control in order to obtain devices with high performance. Such properties are significantly affected by the deposition parameters.
The work presented here focuses on the development and optimisation studies which have been carried out in order to obtain a suitable thin NbTiN film for the development of KI-TWPA devices. The optimised film has been patterned into a KI-TWPA prototype based on an artificial transmission line, which has shown promising performance.

\section{Development of the microfabrication process}
\label{SEC:Fabrication}

The non-linearity required to achieve parametric amplification in the transmission of a wave through a medium can be achieved exploiting the non-linear inductance of Josephson junctions or the non-linear kinetic inductance of superconducting wires ~\cite{Parmenter1962}.
The latter is typically rather weak and, therefore, the challenge of producing kinetic-inductance-based travelling-wave parametric amplifiers with sufficient gain boils down to the challenge of fabricating a sufficiently non-linear low-dissipative medium with a realisable length. 
Once the deposition process to achieve a superconducting film with the required non-linearity is optimised, the remaining microfabrication steps to produce the final KI-TWPA devices are relatively simple, as described in section \ref{SUBSEC:Fab_KI-TWPAs}.

\subsection{Choice of the material}

In order to achieve parametric amplification, the non-linearity of the kinetic inductance of superconducting films can be exploited.
The kinetic inductance is expected to vary non-linearly with the current $I$, in a quadratic fashion at the lowest order: $L_\mathrm{k}(I) \approx L_0\cdot(1 + I^2 / I_*^2)$, where $L_0 = L_\mathrm{k}(I=0)$, $I_*$ sets the scale of the non-linearity and is proportional to $1/\sqrt{R_\mathrm{n}}$, where $R_\mathrm{n}$ is the normal resistance ~\cite{Zmuidzinas2012}.
Thus, high-resistivity superconductors, such as TiN or NbTiN are the ideal candidates to obtain a feasible travelling wave geometry. 
In fact, TiN and NbTiN superconducting films can reach normal resistivity values in the order of $\sim 100 \, \mathrm{\upmu \Omega cm}$ ~\cite{Eom2012}, thus orders of magnitude larger than for aluminium films.

According to the Mattis-Bardeen theory, the linear contribution to the sheet kinetic inductance $L_0$ of a superconducting wire with normal sheet resistance $R_\mathrm{S}$ and gap parameter $\Delta$ is given by~~\cite{MattisBardeen1958}: $L_0 = \hbar R_\mathrm{S} / \pi \Delta$.
In the theoretical framework of the BCS theory~~\cite{Bardeen1957}, the gap parameter $\Delta$ is proportional to the critical temperature $T_\mathrm{c}$ of the superconductor, according to:

\begin{equation} \label{EQ:energy_gap_Tc}
    \Delta = T_\mathrm{c} \cdot k_\mathrm{B} \cdot 1.762
\end{equation}

\noindent where $k_\mathrm{B}$ is the Boltzmann constant. This, in turn, leads to the relation:

\begin{equation} \label{EQ:Lk}
    L_0 = \frac{\hbar \cdot R_\mathrm{S}}{\pi \cdot T_\mathrm{c} \cdot k_\mathrm{B} \cdot 1.762} \, .
\end{equation}

Therefore, a good control on the resistivity (or sheet resistance) and on the critical temperature of the film is a fundamental ingredient in the development of the microfabrication processes for such devices.
A high critical temperature of the film is preferable to allow higher currents during the operation of the amplifiers. Moreover, in superconducting quantum circuits, the number of thermal quasi-particles is reduced if the operational temperature is much lower than the critical temperature. Thus, assuming that the typical operational temperature in a dilution refrigerator is fixed at about $20 \, \mathrm{mK}$, higher critical temperatures, in the order of 10$\,$K, allow to reduce the losses due to quasi-particles \cite{McRae2020}.

NbTiN is currently one of the most promising materials for the development of state-of-the-art KI-TWPAs, with a transition temperature of about 13$\,$K and a kinetic inductance of about 10$\,$pH/sq ~\cite{Eom2012,Malnou2021}.
Our current KI-TWPA prototype design is optimised for a kinetic inductance of 7-10$\,$pH/sq, which guarantees impedance matching to 50$\, \Omega$.
Higher kinetic inductance values could be beneficial to reach higher non-linearity and higher gain, but they would require longer interdigit capacitors, making the geometry much less compact and the fabrication yield lower. More in general, the targeted kinetic inductance range represents a trade-off between different requirements, including lithography resolution, control on film thickness, interdigit capacitor length and line dispersion. \\
The studies performed in this work are aimed at correlating the properties of a sputtered-deposited thin NbTiN film to the sputtering parameters, gaining control on the film characteristics and targeting a kinetic inductance in the range $(7 - 10)\,$pH/sq as well as a critical temperature in the range $(12 - 13) \, \mathrm{K}$.

\subsection{Development of the deposition process}

The deposition process of NbTiN films makes use of rf-magnetron sputter deposition in an atmosphere of argon, using a Nb$_{80\%}$Ti$_{20\%}$ target and introducing nitrogen in the chamber during the process. Sputter deposition allows high-purity films, while other methods such as Atomic Layer Deposition of Chemical Vapour Deposition inevitably suffer from the presence of residual chemical impurities from the organic precursors.

The sputtering system features a load-lock chamber and two separate deposition chambers with four targets each and it is characterised by bottom-up sputtering\footnote{Sputtering system: Kenosistec Physical Vapor Deposition, model KS 800 C Cluster} . The vacuum in the deposition chambers reaches values in the order of $10^{-8} \, \mathrm{mbar}$.
As substrate, high-resistivity 6'' silicon wafers\footnote{FZ silicon wafers, diameter: 6", thickness: $625 \pm 15  \upmu \mathrm{m}$, dopant type: p, dopant: Boron, orientation: $<$100$>$ resistivity: $> 8000 \, \Omega / \mathrm{cm}$} with a $\mathrm{SiO_2}$ layer of about $260 \, \mathrm{nm}$ have been used.

The growing process of superconducting thin films prepared by sputtering is significantly influenced by the sputtering rate and the sputtering temperature, which affect the mobility of the sputtered atoms in the substrate plane ~\cite{Cizek_2006}.
In turn, the power, the pressure as well as the ratio of the nitrogen flow and the argon flow affect the sputtering rate ~\cite{Iosad2003}. In particular, higher power, higher pressure and higher nitrogen flow accelerate the deposition rate. 
Furthermore, the chuck temperature can influence the granularity, the density of defects and the mechanical stress of the film, with direct effects on the critical temperature and the normal resistivity ~\cite{LeiYu2005}.
Thus, the relevant parameters of the sputter deposition process are the cathode power, the chamber pressure, the argon flow, the nitrogen flow, the temperature of the chuck where the wafer is placed and the deposition time. 

In order to reduce the parameter landscape to be explored, the chuck temperature was fixed at $T = 400 \,$\textdegree C (except for one test at $T = 300 \,$\textdegree C), since this value proved to be optimal for niobium films deposition in the same sputtering system, allowing larger granular size and more defined crystalline structures. Before starting the deposition, the substrate is kept at this temperature for a few minutes to allow outgassing.
The test at 300 \textdegree C is useful to verify that the sheet resistance increases with lower deposition temperatures, as expected because of the higher density of defects \cite{Cizek_2006}. 

The first part of the deposition studies (fabrication run \textit{R1}) is dedicated to the investigation of the relevant process parameters.
The second part of the studies (fabrication run \textit{R2}) is instead focused on the calibration of the sputtering rate with a fixed deposition recipe and on the analysis of the variation of the critical temperature $T_\mathrm{c}$ and of the normal sheet resistance $R_\mathrm{s}$ as a function of the film thickness.

\paragraph{Fabrication run \textit{R1}}
In order to explore the parameter landscape of the deposition process, in the first fabrication run \textit{R1} the cathode power $P$, the pressure $p$ and the ratio of the the argon flow and the nitrogen flow $f_\mathrm{Ar}/f_\mathrm{N_2}$ have been varied, as shown in table \ref{TAB:runR1}. The argon flux was maintained at $f_\mathrm{Ar} = 50 \, \mathrm{sccm}$ (and only the nitrogen flux was varied), the chuck temperature at $T = 400 \,$\textdegree C (except for one test at $T = 300 \,$\textdegree C) and the deposition time at $t = 6$ minutes. No photo-lithography and structuring steps were performed during this fabrication run.

\begin{table}[!t] 
\begin{center}
\begin{tabular}{ |p{0.8cm}||p{0.8cm}|p{1cm}|p{1.3cm}|p{1.3cm}|p{0.7cm}|p{1.4cm}|p{1.8cm}|  }
 \hline
 \multicolumn{8}{|c|}{Fabrication run \textit{R1}} \\
 \hline
 Wafer&$P$/W & $p$/mbar & $f_\mathrm{Ar}$/sccm & $f_\mathrm{N_2}$/sccm & $T_\mathrm{c}$/K & $R_\mathrm{s}$/$\Omega$\,sq$^{-1}$ & $L_0$/pH\,sq$^{-1}$\\
 \hline
 T1 & 700 & 2e-3 & 50 & 5 & 10.16 & 17.88 & 2.43\\
 T2 & 700 & 3e-3 & 50 & 4 & 5.15 & 21.5 & 5.77\\
 T3 & 700 & 3e-3 & 50 & 5 & 10.38 & 17.75 & 2.36\\
 T4 & 700 & 3e-3 & 50 & 6 & 13.9 & 15.63 & 1.55\\
 T5 & 1200 & 3e-3 & 50 & 5 & $<$4.2 & - & -\\
 T6 & 700 & 3e-3 & 50 & 7 & 14.17 & 19.63 & 1.91\\
 T7 & 700 & 3e-3 & 50 & 8 & 13.16 & 29.75 & 3.12\\
 T8 & 700 & 3e-3 & 50 & 6.5 & 13.86 & 19.25 & 1.91\\
 T9* & 700 & 3e-3 & 50 & 7 & 13.78 & 21.13 & 2.12\\
 T10 & 600 & 3e-3 & 50 & 7 & 13.76 & 27.75 & 2.79\\
 \hline
\end{tabular}
\end{center}
*$T = 300 \,$\textdegree C
\caption{Sputtering parameters that were varied during the fabrication run \textit{R1}. Argon flux, chuck temperature and time were instead fixed respectively at the following values: $f_\mathrm{Ar} = 50 \, \mathrm{sccm}$, $T = 400 \,$\textdegree C (except for one test at $T = 300 \,$\textdegree C marked with *), $t = 6$ minutes. The measured critical temperature $T_\mathrm{c}$, the normal sheet resistance at low temperature, measured before the transition, $R_\mathrm{s}$ and the resulting estimated kinetic inductance $L_\mathrm{k}(0)$ are also reported.}
\label{TAB:runR1}
\end{table}

\begin{figure}[t!]
    \centering
    \begin{subfigure}[t]{0.32\textwidth}
        \centering
        \includegraphics[height=4.2cm, trim={1cm 0 0.8cm 0}]{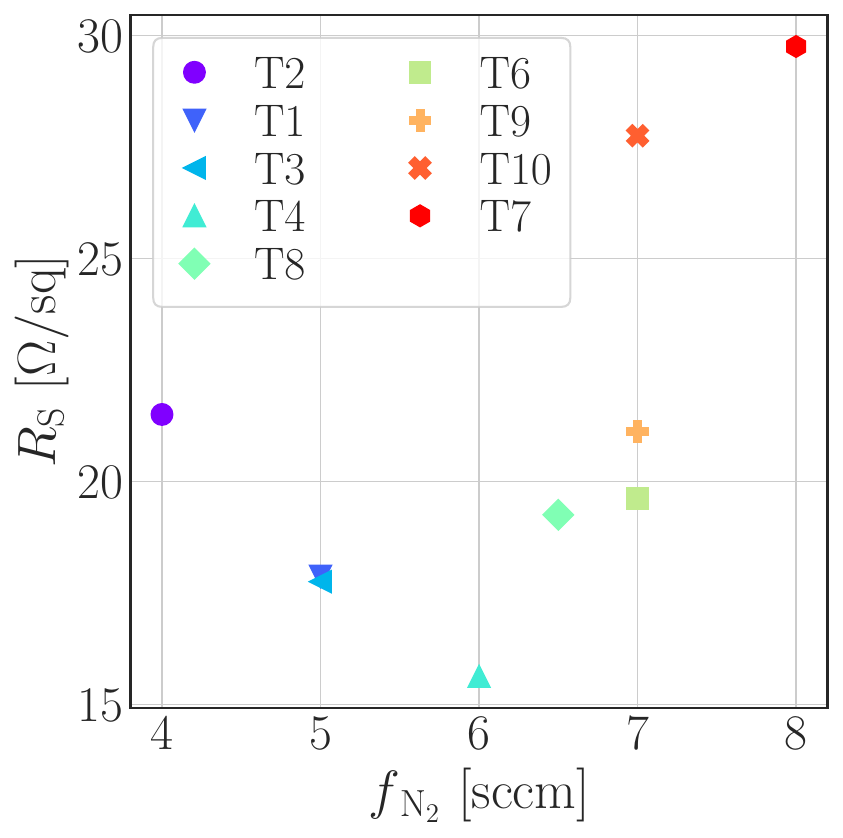}
        \caption{} \label{SUBFIG:DWT2_Rs_vs_fN2}
    \end{subfigure}
    \begin{subfigure}[t]{0.32\textwidth}
        \centering
        \includegraphics[height=4.2cm, trim={1cm 0 0.8cm 0}]{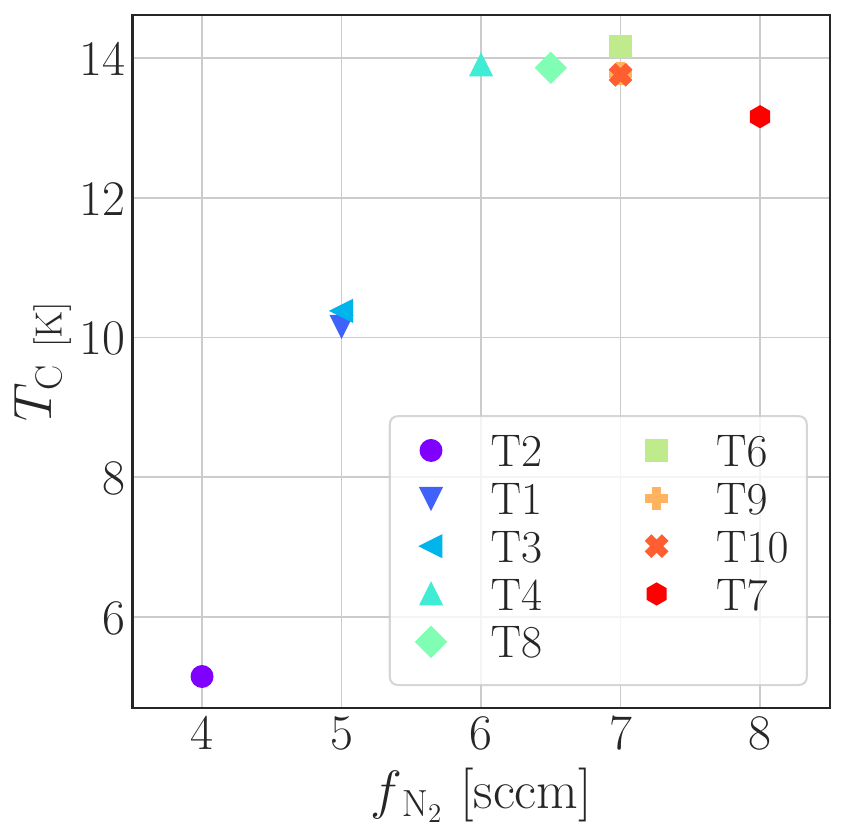}
        \caption{} \label{SUBFIG:DWT2_Tc_vs_fN2}
    \end{subfigure}
    \begin{subfigure}[t]{0.32\textwidth}
        \centering
        \includegraphics[height=4.2cm, trim={1cm 0 0.8cm 0}]{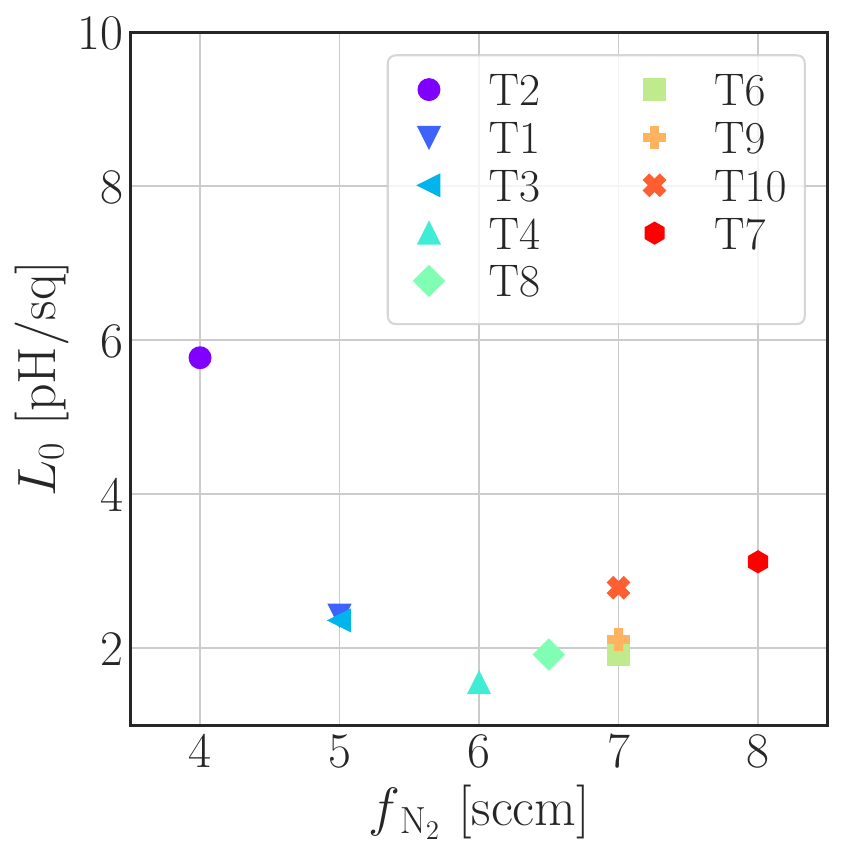}
        \caption{} \label{SUBFIG:DWT2_Lk_vs_fN2}
    \end{subfigure}

    \caption{Experimental parameters, sheet resistance in a), critical temperature in b) and kinetic inductance in c), as a function of nitrogen flow, for the different recipes tested in fabrication run \textit{R1}, as reported in table \ref{TAB:runR1}.}
    \label{FIG:DWT2_plots}
\end{figure}

After the sputter deposition, the fabricated wafers were diced in samples with a size of $17 \, \mathrm{mm} \times 2 \, \mathrm{mm}$ and characterised at $4.2 \, \mathrm{K}$ in a liquid helium bath.
The superconducting transition of each sample has been measured, extrapolating the critical temperature $T_\mathrm{c}$ and the resistance before the transition $R_\mathrm{N}$, from which the normal sheet resistance $R_\mathrm{s}$ was calculated. Finally, the linear term of the kinetic inductance $L_0$ was estimated, according to equation \ref{EQ:Lk}. The results of these measurements are reported in table \ref{TAB:runR1} and shown in figure \ref{FIG:DWT2_plots}.

The thickness of the resulting film was not equal for the different samples, as the sputtering rate is influenced by the power and the pressure, for example, and the deposition time was fixed.
Both the normal sheet resistance and the critical temperature, even if the latter to a lesser degree, are expected to be dependent on the film thickness.
However, the measurement of these two parameters for the different samples recipes proved to be a useful tool to pinpoint the most promising deposition recipe, to be then investigated in more detail and fine-tuned in the next fabrication run, \textit{R2}.

\paragraph{Fabrication run \textit{R2}}

Among the recipes tested in the fabrication run \textit{R1}, the recipe used for wafer T10 with a power of $P =600 \, \mathrm{W}$, a pressure of $p = 3 \times 10^{-3} \, \mathrm{mbar}$, a ratio of argon flow and nitrogen flow of $f_\mathrm{Ar}/f_\mathrm{N_2} = 50/7$ and a chuck temperature of $T = 400 \,$\textdegree C has been chosen as a basis for the next tests. In fact, the resulting sheet resistance achieved with this recipe was rather high (about $28 \, \Omega/\mathrm{sq}$) if compared with the other results, keeping, however, the critical temperature sufficiently high (about $13.8 \, \mathrm{K}$). The calculated kinetic inductance is about $2.1 \, \mathrm{pH/sq}$, thus far below the target range of $(7 - 10)\,$pH/sq. However, as previously mentioned, both the critical temperature and the sheet resistance, which in turn define the kinetic inductance of the film according to theory, depend on the thickness of the film and, therefore, they can be fine tuned varying the deposition time.
The second fabrication run, \textit{R2}, is indeed dedicated to the investigation of the variation of these parameters as a function of the film thickness. Six wafers have been sputter deposited with a fixed recipe (run \textit{R1}, recipe used for wafer T10) and varying the deposition time, as reported in table \ref{TAB:runR2}.

\begin{table}[t!]
\centering
\begin{tabular}{ |p{0.8cm}||p{1.4cm}|p{1.27cm}|p{0.8cm}|p{1.5cm}|p{1.9cm}|  }
 \hline
 \multicolumn{6}{|c|}{Fabrication run \textit{R2}} \\
 \hline
 Wafer & $t$/minutes & $T$/nm & $T_\mathrm{c}$/K & $R_\mathrm{s}$/$\Omega$\,sq$^{-1}$ & $L_0$/pH\,sq$^{-1}$\\
 \hline
 W1 & 6.0 & 29.8   & 13.45 & 32.4 & 3.32\\
 W2 & 2.0 & 7.4    & 12.03 & 125 & 14.7\\
 W3 & 9.0 & 43.8   & 14.07 & 18 & 1.75\\
 W4 & 4.0 & 19.5   & 13.35 & 44 & 4.70\\
 W5 & 3.0 & 14.5   & 12.88 & 62 & 6.64\\
 W6 & 2.5 & 10.8 & 12.45 & 86 & 9.53\\
 \hline
\end{tabular}
\caption{Deposition times $t$ and the corresponding measured thickness values $T$. The measured critical temperature $T_\mathrm{c}$, the normal sheet resistance at low temperature, measured before the transition, $R_\mathrm{s}$ and the resulting estimated kinetic inductance $L_0$ are also reported.}
\label{TAB:runR2}
\end{table}

\begin{figure}[t!]
    \centering
    \includegraphics[width=0.6\linewidth]{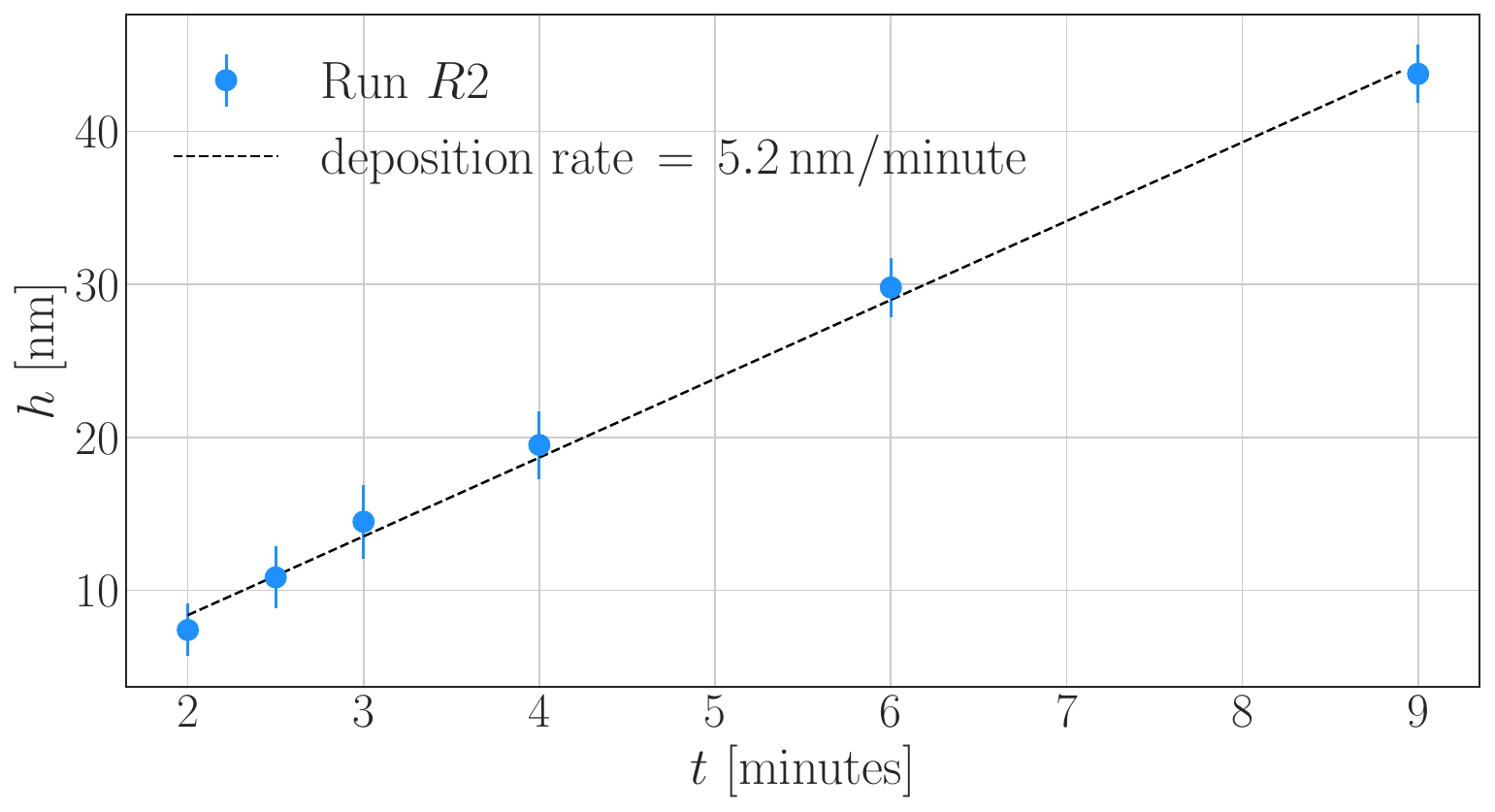}
    \caption{Measured thickness as a function of the deposition time for the wafers of fabrication run \textit{R2}. The data are fit with a linear function which returns the deposition rate of the sputtering process, as reported in the legend.}
    \label{FIG:DWT3_deposition_rate}
\end{figure}

In order to precisely measure the thickness of the deposited NbTiN film, a lithographic step followed by an etching step have been introduced. In this way, it is possible to directly measure the step height from the etched areas - where only the oxide layer is left - and the NbTiN film. However, during the ICP-RIE process a more or less pronounced overetch is unavoidable, leading to a reduced thickness of the oxide in the etched areas with respect to the not-etched areas. Thus, a measurement of the oxide thickness before and after the etching process is necessary to correct for this. The oxide thickness has been measured with an optical interferometer\footnote{Interferometer LEITZ Ergolux}, while the step from the oxide layer in the etched areas to the NbTiN film has been measured with an atomic force microscope\footnote{AFM Px Nt-Mdt PX}.
Figure \ref{FIG:DWT3_deposition_rate} shows the estimated film thickness values as a function of the sputtering time, leading to a sputtering rate of $5.2 \, \mathrm{nm/minute}$.

\begin{figure}[t!]
    \centering
    \begin{subfigure}[t]{0.32\textwidth}
        \centering
        \includegraphics[height=4.15cm, trim={0.5cm 0 0.5cm 0}]{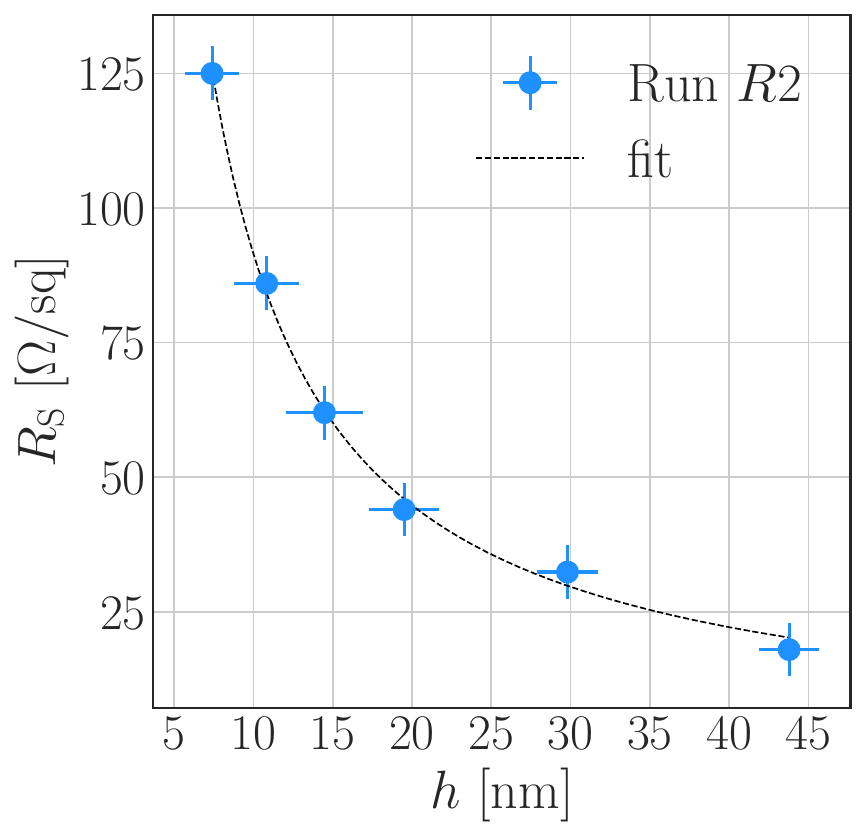}
        \caption{} \label{SUBFIG:DWT3_Rs_vs_thickness}
    \end{subfigure}
    \begin{subfigure}[t]{0.32\textwidth}
        \centering
        \includegraphics[height=4.15cm, trim={0.5cm 0 0.5cm 0}]{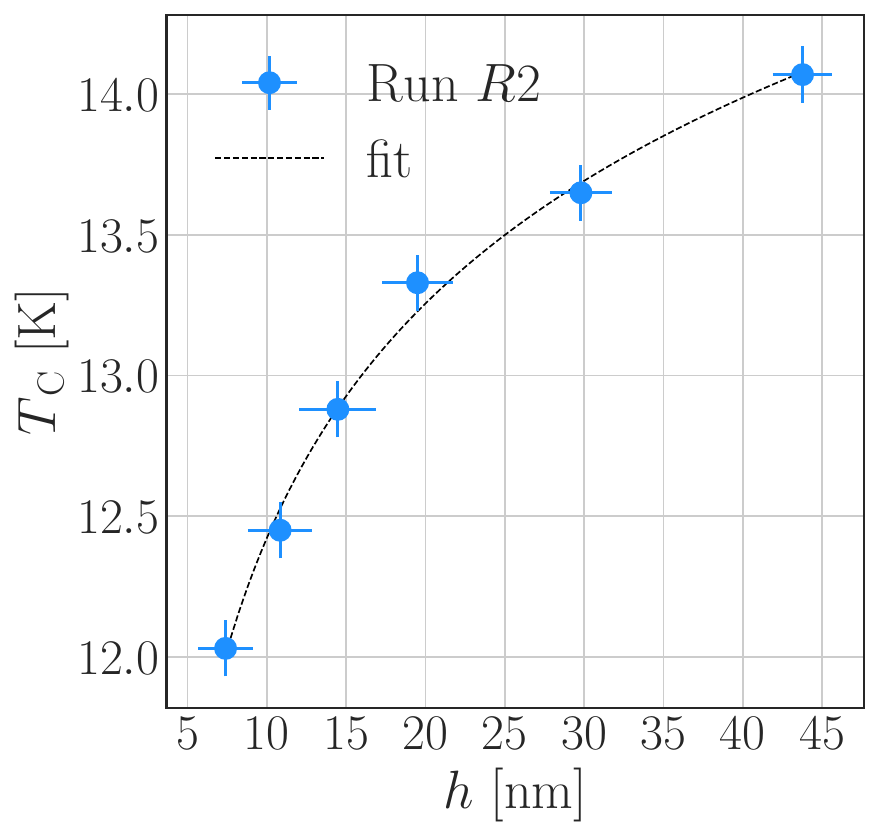}
        \caption{} \label{SUBFIG:DWT3_Tc_vs_thickness}
    \end{subfigure}
    \begin{subfigure}[t]{0.32\textwidth}
        \centering
        \includegraphics[height=4.15cm, trim={0.5cm 0 0.5cm 0}]{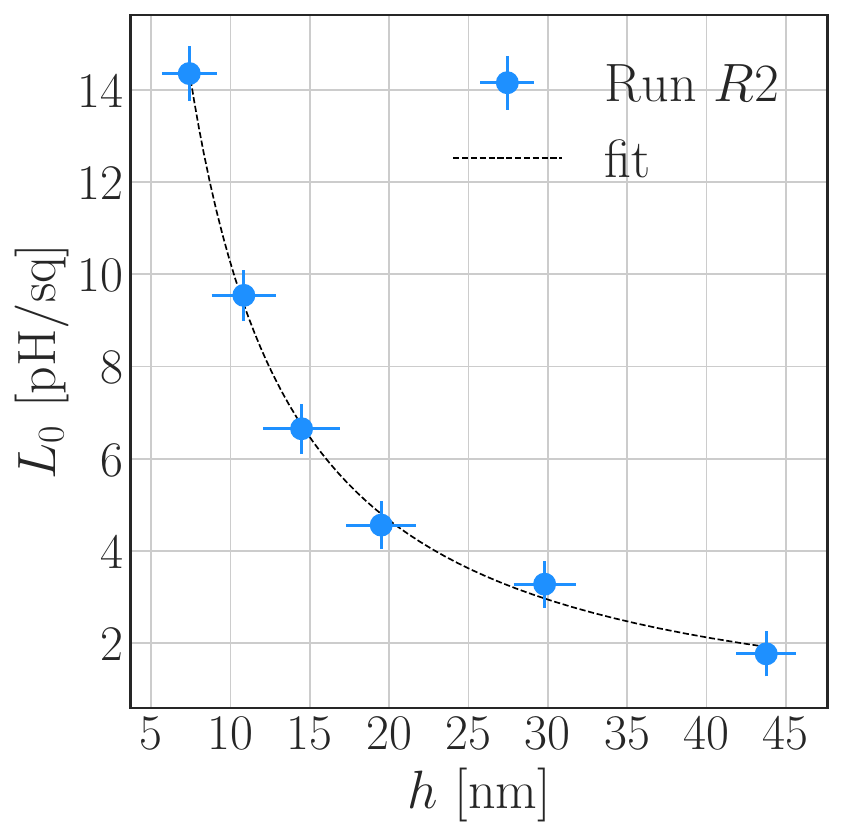}
        \caption{} \label{SUBFIG:DWT3_Lk_vs_thickness}
    \end{subfigure}

    \caption{Experimental parameters, sheet resistance in a), critical temperature in b) and kinetic inductance in c), as a function of thickness, for the wafers of fabrication run \textit{R2}, as reported in table \ref{TAB:runR2}. The sheet resistance data are fit with Fuchs' model ~\cite{fuchs_1938}, while the critical temperature values are fit with a scaling model \cite{Ivry2014} and the kinetic inductance values are fit with a phenomenological model \cite{Zmuidzinas2012}, as explained in the text.}
    
    \label{FIG:DWT3_plots}
\end{figure}

\begin{figure}[t!]
	\centering
	\includegraphics[width=0.6\textwidth]{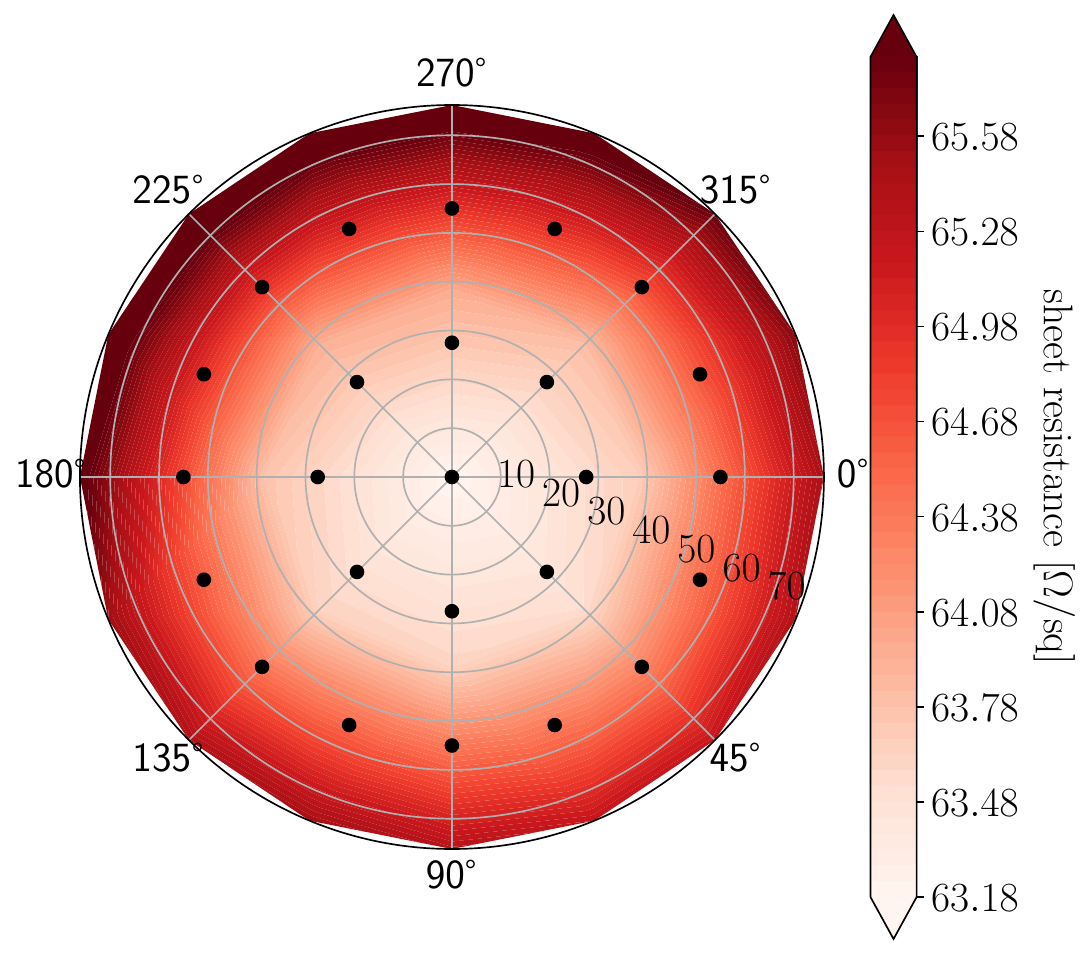}
	\caption{Example of the sheet resistance variations measured on the wafer W6 of run \textit{R2}. The percentage variation is about $4\%$. The 25 black dots show the position of the measurement points.} 
	\label{FIG:sheet_res_colormap}
\end{figure}

The kinetic inductance of the film was calculated from the measured normal sheet resistance and the measured critical temperature analogously to run \textit{R1}.
To characterise the film properties at $4.2 \, \mathrm{K}$, a chip with a $1 \, \mathrm{mm}$ wide strip-like structure patterned into the film has been employed.
As expected, both the normal sheet resistance and the critical temperature show a dependence on the film thickness, as visible in figure \ref{FIG:DWT3_plots}.
According to Fuchs' model ~\cite{fuchs_1938}, the sheet resistance $R_\mathrm{S}$ should scale with the film thickness $h$ as $ R_\mathrm{S}(h) \sim \rho/h \cdot (1+c/h)$, where $\rho$ and $c$ are parameters of the model. In this model, the first parameter, $\rho = 87 \, \upmu \Omega \mathrm{cm}$, corresponds to the resistivity of the bulk material. The experimental data is shown in figure \ref{SUBFIG:DWT3_Rs_vs_thickness}, together with a fit with the theoretical model. 
Figure \ref{SUBFIG:DWT3_Tc_vs_thickness} shows the critical temperature $T_\mathrm{C}$ as a function of the estimated film thickness $h$. The data is fit with the scaling model $T_\mathrm{C}(h) \propto 1/h \cdot R_\mathrm{S}^{-k}$, where $k=1.06$ is a free parameter returned from the fit, which results close to one, as expected \cite{Ivry2014}. 
The resulting kinetic inductance, calculated according to equation \ref{EQ:Lk}, is shown in figure \ref{SUBFIG:DWT3_Lk_vs_thickness}. The data is fit with the phenomenological model $L_0(h) \propto 1/h^\alpha$, where $h$ is the film thickness and $\alpha = 1.13$ is a free parameter in the fit ~\cite{Zmuidzinas2012}.
For the film with an estimated thickness of about $11 \, \mathrm{nm}$ (wafer W6), the resulting kinetic inductance is about $9.5 \, \mathrm{pH/sq}$, within the design range of $(7 - 10)\,$pH/sq.

The uniformity of the film thickness on a wafer scale has a direct impact on the uniformity of the film properties, such as the critical temperature and the normal sheet resistance.
The wafer is placed at a distance of about $200 \, \mathrm{mm}$ from the NbTi cathode, ensuring a sufficient solid angle to cover the wafer surface. The wafer is rotating with a speed of $15 \pm 0.1 \, \mathrm{rpm}$ to increase the uniformity of the deposition process. 
In order to evaluate the uniformity of the film thickness, a measurement of the sheet resistance of each deposited film in 25 points distributed over the areas of the 6'' wafer substrates has been performed\footnote{CRESBOX 4 point probe sheet resistance/resistivity measurement, Nepson Corporation}. The variation of the sheet resistance of the film is expected to be primarily caused by the variation of the film thickness. Thus, the measured percentage variation of the sheet resistance is assumed to be equivalent to the percentage variation of the film thickness. The average measured variation in the six wafers of run \textit{R2} is about 5\%. Figure \ref{FIG:sheet_res_colormap} shows an example of the sheet resistance variations over the wafer W6 (run \textit{R2}). The impact of such level of thickness inhomogeneity on a wafer scale is negligible for the fabrication of the final devices. In fact, a variation of 5\% on the film thickness in the $10 \, \mathrm{nm}$ thickness range translates in an expected variation of the critical temperature of less than 5\%.

\section{Measurement of the kinetic inductance using microwave resonators}

\begin{figure}[t!]
    \centering
    \begin{subfigure}[t]{0.49\textwidth}
        \centering
        \includegraphics[height=6cm, trim={0cm 0 0cm 0}]{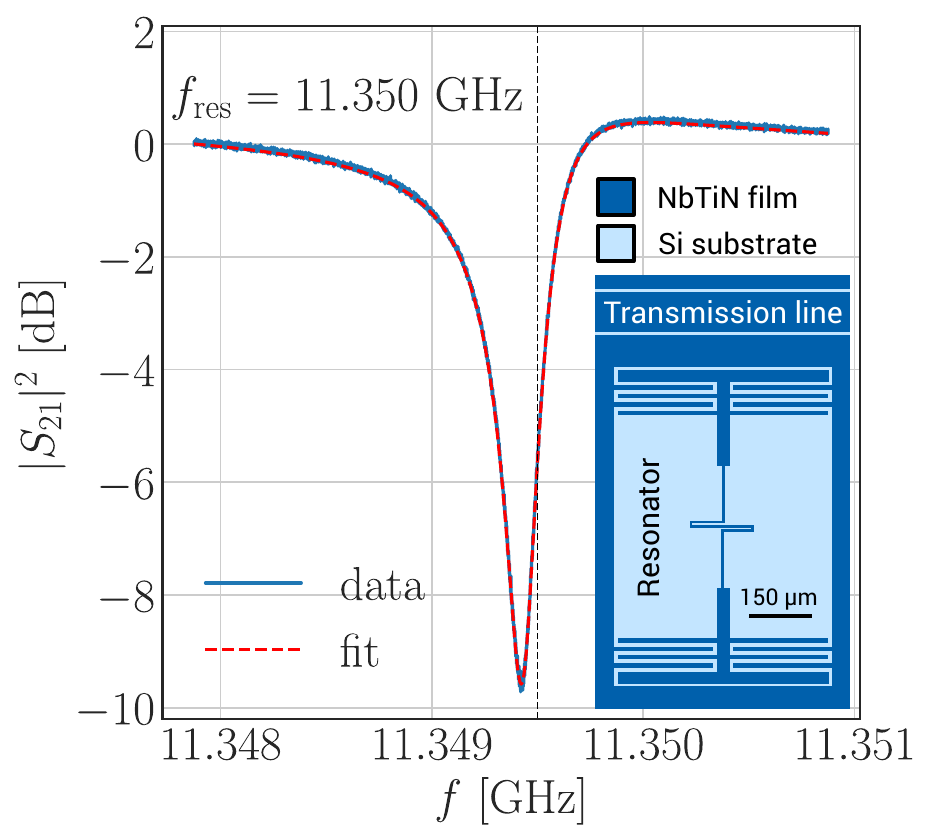}
        \caption{} \label{SUBFIG:Res8_fit}
    \end{subfigure}
    \begin{subfigure}[t]{0.49\textwidth}
        \centering
        \includegraphics[height=6cm, trim={0cm 0 0cm 0}]{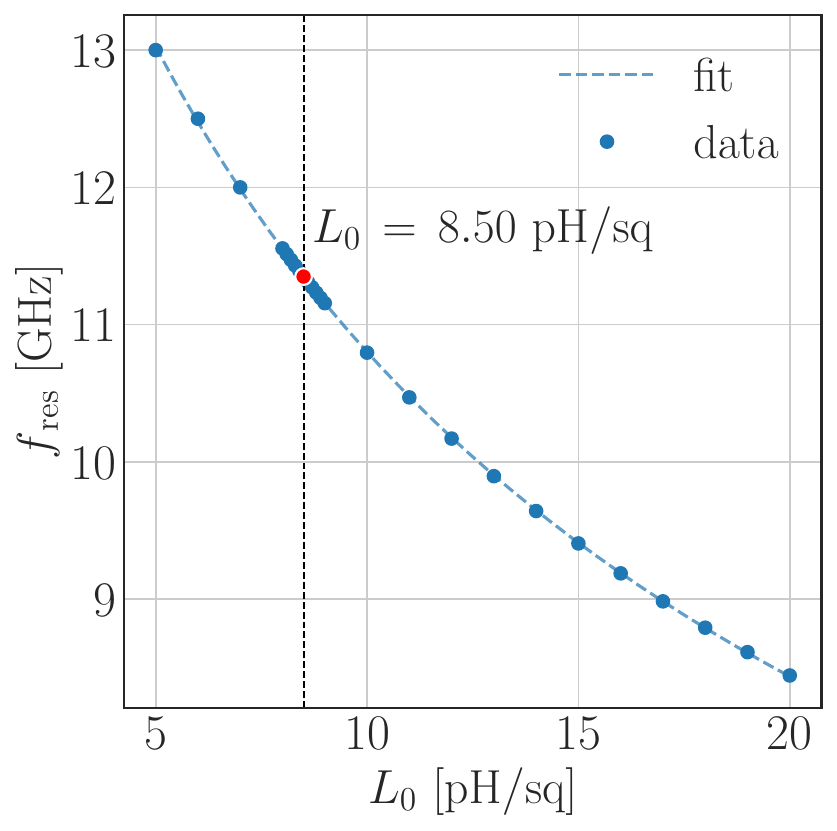}
        \caption{} \label{SUBFIG:Lk_vs_freq}
    \end{subfigure}

    \caption{a) The resonance curve together with the corresponding fit which yields the resonance frequency reported in the figure. The inset shows the design of the lumped element resonator used for this measurement, which is presented in \cite{Mezzena2020}. b) The dependence of the resonance frequency on the kinetic inductance of such a resonator, resulting from dedicated simulations, as explained in the text. The experimental value of the resonance frequency shown in a) yields to the kinetic inductance value reported in the figure.}
    
    \label{FIG:Lk_estimation}
\end{figure}

In order to estimate the value of the kinetic inductance in the fabricated film, lumped element superconducting resonators fabricated with the same recipe of wafer W6 of the fabrication run \textit{R2} were exploited. In this fabrication, high-resistivity 6'' silicon wafers with a thin SiO$_2$ layer of about $40 \, \mathrm{nm}$ have been used. 
The resonance frequency $f_\mathrm{res}$ of such resonators depends on the capacitance $C$, on the geometrical inductance $L_\mathrm{g}$ and on the kinetic inductance $L_\mathrm{k}$, according to the following relation: $f_\mathrm{res} = 1/\sqrt{(L_\mathrm{g}+L_\mathrm{k})\cdot C}$.
Therefore, by simulating\footnote{Simulation software: Sonnet Software} the resonators and assuming different values of the kinetic inductance ~\cite{Wisbey2014}, it is possible to obtain a calibration curve of the kinetic inductance as function of the resonance frequency, as shown in figure \ref{SUBFIG:Lk_vs_freq}.

The measurements of the resonators were performed at millikelvin temperature. The chip used for the measurements consists of eight lumped element resonators with unique resonance frequencies coupled to a common transmission line. 
The measured transmission $S_{21}$ as a function of frequency for one of these resonators and the corresponding fit is shown in figure \ref{SUBFIG:Res8_fit}, yielding to a resonance frequency of $11.350 \, \mathrm{GHz}$.
Comparing this result with the calibration curve, the resulting kinetic inductance is about $8.5 \, \mathrm{pH/sq}$, as shown in figure \ref{SUBFIG:Lk_vs_freq}. This value of the kinetic inductance is shifted by about 10\% with respect to the value reported in table \ref{TAB:runR2}, calculated according to equation \ref{EQ:Lk}.
This discrepancy is within the statistical and systematic uncertainties of the measurements and of the simulation results.
Furthermore, the chip with the measured resonator was fabricated on a different wafer than wafer W6 of the fabrication run \textit{R2} and a certain degree of inhomogeneity is to be expected.
Finally, it has to be considered that the theoretical coefficient of $1.762$~\cite{Bardeen1957} reported in equation \ref{EQ:energy_gap_Tc} can be material dependent and, thus, not fully accurate for NbTiN films~\cite{KOMENOU1968335, Beck.PhysRevLett.107.177007}. 

By applying a standard resonator fitting routine based on a Lorentzian model \cite{GaoPhDThesis} we also extrapolated the internal quality factors of the measured resonators. The obtained values resulted in the $Q_\mathrm{i}=(30 - 200)\cdot 10^3$ range. The attenuation of a transmission line of length $z$ is given by  $I(z)=\exp{[(-2\pi\,\delta\,z)/\lambda]}$~~\cite{Pozar2011}, where $\delta=1/(2Q_\mathrm{i})$~~\cite{Gao2008} is the tangent loss and $\lambda$ is the wavelength of the pump that travels on the line. 
As described in section \ref{SUBSEC:KI-TWPA_design}, the KI-TWPA design consists of an artificial transmission line made up of repeated elementary structures called \textit{super-cells}.
By considering that the length of a \textit{super-cell} is $L_\mathrm{sc}=\lambda/2$ and that $z = L_\mathrm{sc} \cdot N_\mathrm{sc}$ (where $N_\mathrm{sc}$ is the number of \textit{super-cells} in the KI-TWPA line), we can rewrite the attenuation as $I(z)=\exp{[-\pi\,N_\mathrm{sc}/(2Q_i)]}$. This relationship depends on the internal quality factor $Q_\mathrm{i}$ and the number of \textit{super-cells} in the line. Considering about 1000 \textit{super-cells} in the final KI-TWPA design, with the measured internal quality factors the attenuation would be in the range $(0.95-0.99)$. This means that the total insertion loss due to the amplifier falls in the range from -0.45 to -0.09$\,$dB that is negligible if compared with losses due to other components present in the read-out line.

\section{Realisation of a KI-TWPA prototype}
\label{SEC:KI-TWPA_device}

Kinetic inductance-based TWPA (KI-TWPA) have gained interest in the scientific community not only thanks the their potentially wide amplification bandwidth, noise near the quantum limit, and high dynamic range ~\cite{Eom2012,Malnou2021}, but also thanks to their relatively simple fabrication.
In fact, these devices require only one lithography step (patterning of the superconducting film) and one etching step, as described in section \ref{SEC:Fabrication}.
However, the engineering and production of a long non-linear medium requires advanced micro/nano-fabrication methodologies and poses issues such as $50\,\Omega$ impedance matching and phase matching ~\cite{Esposito2021}. 
In order to demonstrate the suitability of the NbTiN film optimised as described in the previous sections, we have exploited such film to produce a KI-TWPA prototype within the DARTWARS project. The prototype underwent a preliminary characterisation, showing promising performance, as described in the following sections.

\subsection{KI-TWPA design}
\label{SUBSEC:KI-TWPA_design}

\begin{figure}[t!]
	\centering
	\includegraphics[width=0.8\textwidth]{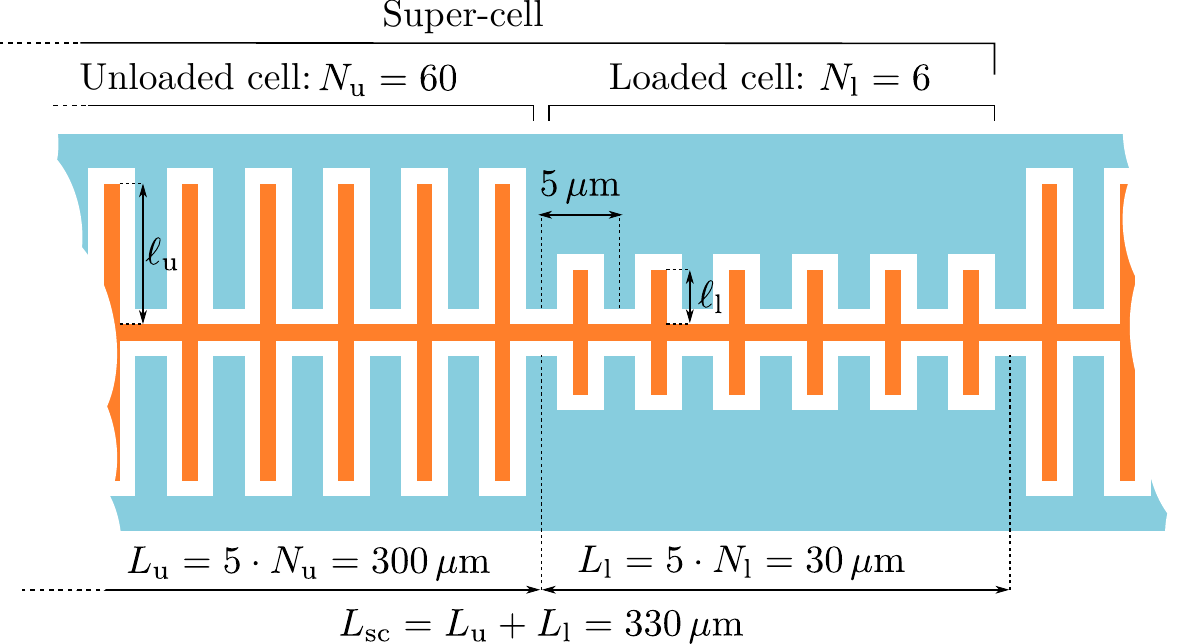}
	\caption{The diagram (not to scale) depicts a \textit{super-cell} composed of $N_\mathrm{u}$ elementary cells, with a finger length of $\ell_\mathrm{u}=102\,\upmu$m and characteristic impedance of $Z_0=50\,\Omega$ (\textit{unloaded cells}), as well as $N_\mathrm{l}$ elementary cells with a finger length of $\ell_\mathrm{l}=33.5\,\upmu$m and characteristic impedance of $Z_0=80\,\Omega$ (\textit{loaded cells}). The \textit{super-cell} is designed such that its length, $L_\mathrm{sc}$, is equal to half of the wavelength of the pump used to operate the parametric amplifier. In its final configuration, the KI-TWPA amplifier will be composed of $N_\mathrm{sc}=1000$ \textit{super-cells}, resulting in a total length of approximately 33$\,$cm. The conducting track of the line is reported in orange while the ground plane is reported in blue. The conducting track of the CPW line is shown in orange, while the ground plane is depicted in blue. The gap between the conducting track and ground (shown in white) measures $1\,\upmu$m.} 
	\label{fig:fishbone}
\end{figure}

A KI-TWPA amplifier consists of a weakly dispersive transmission line for which the phase-matched bandwidth, needed to create the exponential amplification, is controlled by dispersion engineering~~\cite{Chaudhuri2017}. 
Amplifiers can be implemented as CPW~\cite{Malnou2021} and micro-strip~\cite{Shu2021} by exploited a fish-bone shaped solution that allows to tune the characteristic impedance to the desired value by tuning the length of the fingers that form the line itself (figure \ref{fig:fishbone}).

The design selected for the DARTWARS is similar to the one proposed in M. Malnou {\it et al.}~\cite{Malnou2021}. The KI-TWPA is composed of a long transmission line made up of multiple elementary cells. Each cell is constructed using a CPW that includes the line inductance $L$, that depends on the material kinetic inductance, and two interdigitated capacitor (IDC) fingers located on either side of the center line, which create the capacitance to ground $C$ (figure \ref{fig:fishbone}). The finger length $\ell$ is adjusted to achieve a $C$ value such that $Z_0 = \sqrt{L/C} = 50\,\Omega$ (\textit{unloaded cells}) as described in reference~\cite{Chaudhuri2017}. To create the exponential parametric amplification, the design employs dispersion engineering by introducing periodic loadings through the reduction of the fingers lengths (\textit{loaded cells}), as explained in references~~\cite{Chaudhuri2017} and~~\cite{Malnou2021}. To achieve an exponential gain in the $4-8\,$GHz range, each \textit{super-cell} is composed by $N_\mathrm{u}=60$ \textit{unloaded cells} and by $N_\mathrm{l}=6$ (\textit{loaded cells}). The finger length for the \textit{loaded cells} is tuned in order to have a characteristic impedance of $Z_0 = \sqrt{L/C} = 80\,\Omega$. Each \textit{super-cell} has a length of $330\,\upmu$m, which is half of the wavelength ($\lambda/2$) of the pump signal in the  $(8 - 12)\,$GHz frequency range. 

In its final configuration, the KI-TWPA amplifier will be composed of $N_\mathrm{sc} = 1000$ \textit{super-cells}, resulting in a total length of approximately 33$\,$cm, for a total gain of around 20$\,$dB.
As first step, to test the optimised NbTiN film and to develop the fabrication process, we first produced a medium-sized device composed of 523 \textit{super-cells}  with a length of about 17$\,$cm. The expected gain for this prototype is in the $(7-11)\,$dB range, depending on the used pump power. While these performance levels may not be competitive, the results from these preliminary amplifiers will be crucial in refining the final design.

\subsection{Microfabrication}
\label{SUBSEC:Fab_KI-TWPAs}

\begin{figure}[t!]
	\centering
	\includegraphics[width=0.75\textwidth]{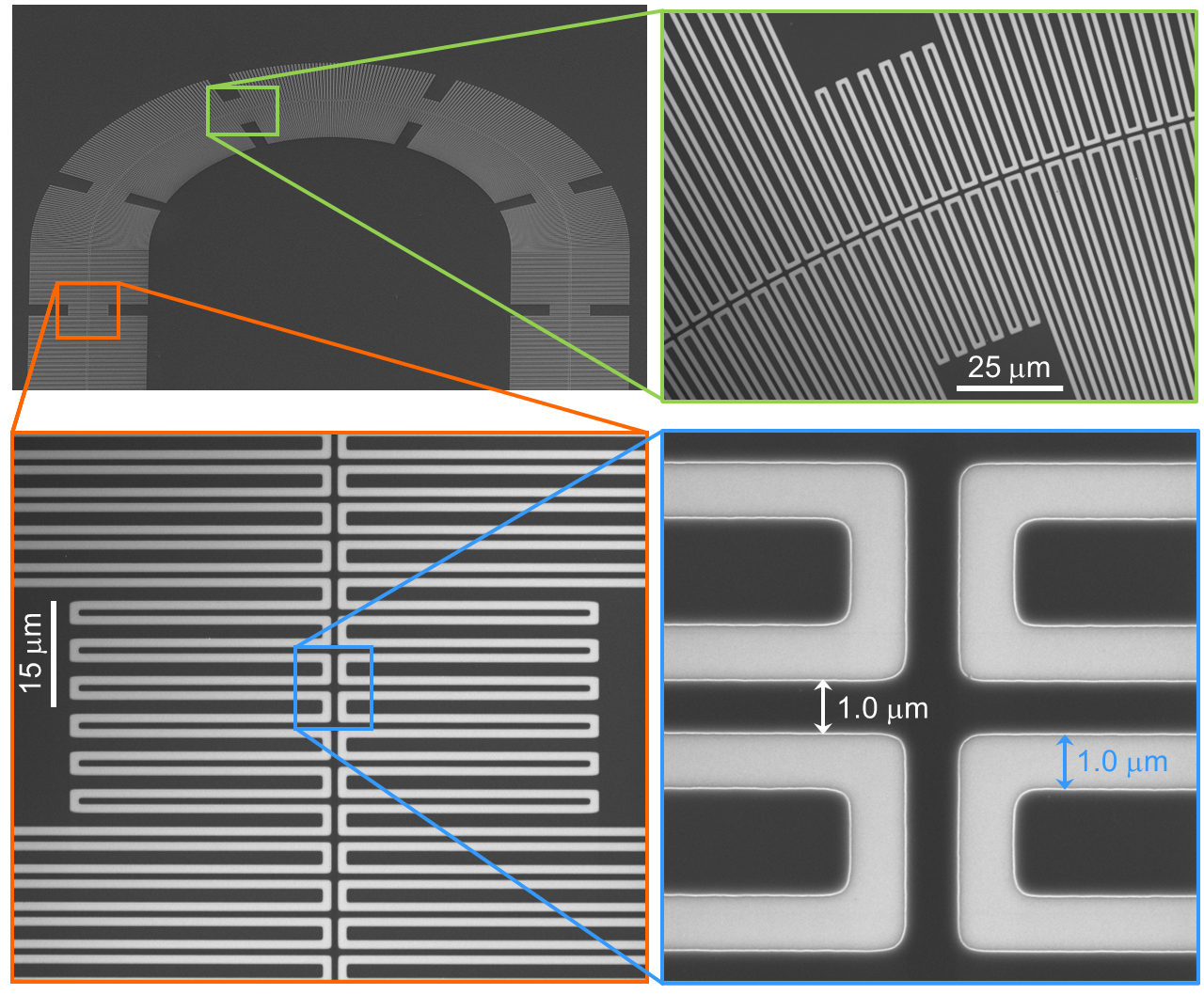}
	\caption{SEM pictures of the fabricated KI-TWPA. In the first image the sample is tilted. The dark (light) areas correspond to the metal (substrate). The design consists of a 14$\,$cm long transmission line with the fish-bone geometry illustrated in figure \ref{fig:fishbone}. The fingers are $1 \, \upmu \mathrm{m}$ wide and $102\, \upmu \mathrm{m}$ ($33.5 \, \upmu \mathrm{m}$) long in the \textit{unloaded cells} (\textit{loaded cells}).} 
	\label{FIG:SEM_photos}
\end{figure}

For the microfabrication of the first KI-TWPA prototype devices, the deposition recipe of wafer W6 of fabrication run \textit{R2} with power $P = 600 \, \mathrm{W}$, pressure $p = 3 \times 10^{-3} \, \mathrm{mbar}$, argon and nitrogen flows $f_\mathrm{Ar}/f_\mathrm{N_2} = 50/7$, chuck temperature $T = 400 \,$\textdegree C and time $t = 2.5 \, \mathrm{minutes}$ was used. Using this recipe, the NbTiN film is deposited on high-resistivity 6'' wafers\footnote{FZ silicon wafers, diameter: 6'', thickness: $625 \pm 15  \upmu \mathrm{m}$, dopant: Boron (type p), orientation: $<$100$>$ resistivity: $> 8000 \, \Omega \mathrm{cm}$} with a thin SiO$_2$ layer on top of about $40 \, \mathrm{nm}$. After that, the photoresist\footnote{Stepper resist OIR 674-09 Fujifilm, thickness of $0.75 \, \mathrm{\upmu m}$} is structured with the geometry described in the previous section. Subsequently, the wafers undergo a Reactive Ion Etching step (Sulfur hexafluoride gas ($\mathrm{SF}_6$), pressure of $150 \, \mathrm{mT}$, power of $150 \, \mathrm{W}$, etching time of $30 \, \mathrm{s}$). Finally, the resist is removed using an acetone bath followed by an isopropyl bath.
Figure \ref{FIG:SEM_photos} shows Scanning Electron Microscope (SEM) pictures of the fabricated devices, with a zoom on a segment of the fish-bone.

\subsection{Preliminary characterisation}
\label{SUBSEC:Characterisation_KI-TWPAs}

\begin{figure}[t!]
    \centering
    \begin{subfigure}[t]{0.49\textwidth}
        \centering
        \includegraphics[height=6cm, trim={0.7cm 0 0.5cm 0}]{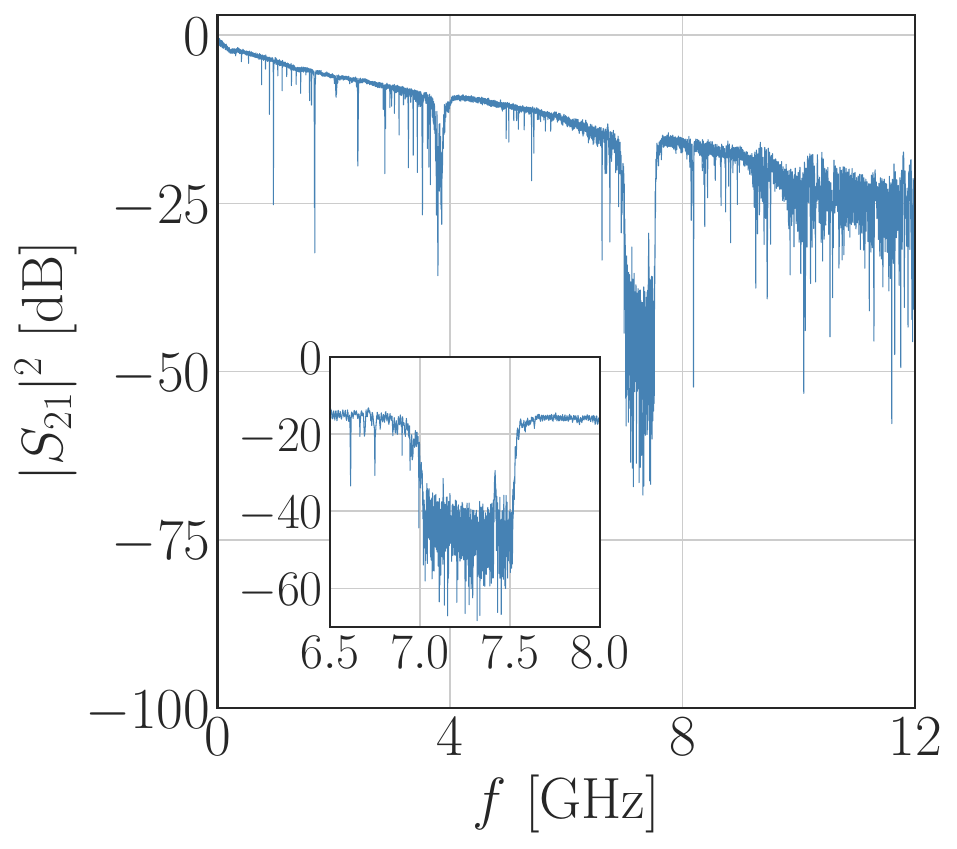}
        \caption{} \label{SUBFIG:DWD2_S21}
    \end{subfigure}
    \begin{subfigure}[t]{0.49\textwidth}
        \centering
        \includegraphics[height=6cm, trim={0.7cm 0 0.5cm 0}]{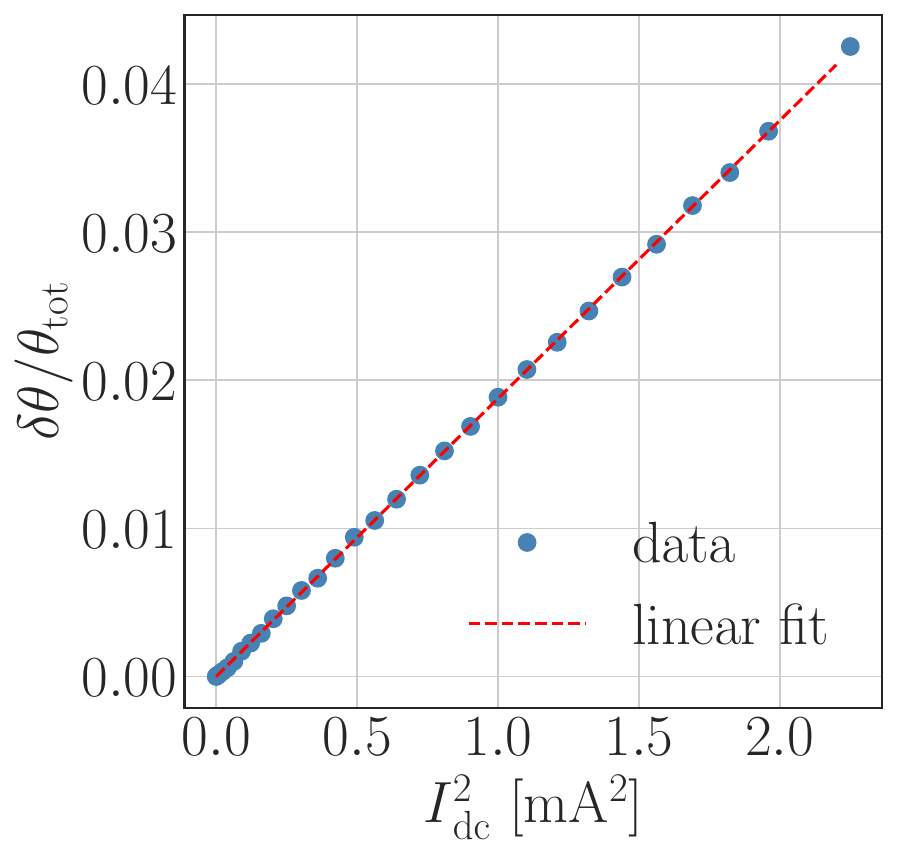}
        \caption{} \label{SUBFIG:DWD2_phase_shift}
    \end{subfigure}

    \caption{(a) Forward transmission parameter $S_{21}$ of the tested KI-TWPA prototype. In the inset the stop band centred at about $7 \, \mathrm{GHz}$ is shown. The band-gap, also known as the stop-band, is a result of the dispersion engineering discussed in Section \ref{SUBSEC:KI-TWPA_design}. (b) Non-linear phase shift as function of the dc current with a quadratic fit.} 
    
    \label{FIG:DWD2_plots}
\end{figure}

The prototype devices underwent a preliminary characterisation at low temperatures with the goal of demonstrating the suitability of the NbTiN films for such application.
The first tests have been conducted at $4.2 \, \mathrm{K}$ by means of a liquid helium bath, using a simplified set-up comprising two coaxial cables and no attenuators or amplifiers. The transmission response of the KI-TWPA prototype has been measured by means of a commercial Vector Network Analyzer (VNA). The corresponding plot is shown in figure \ref{SUBFIG:DWD2_S21}. Due to the weakly dispersive nature of the line, the dispersion engineering discussed in section \ref{SUBSEC:KI-TWPA_design} created a band-gap in the frequency range of 7-7.5$\,$GHz, which is slightly lower than expected. This is not entirely unexpected, as the goal of this first production was to characterise the material and the fabrication processes of the devices, and no optimisations have been performed on the finger length for the \textit{loaded} and \textit{unloaded cells}. 
The spurious resonance at about $4 \, \mathrm{GHz}$ is most likely due to spurious modes in the ground plane of the chip, which are expected, since the packaging has not been optimised yet. 

The critical current of the devices has been measured by superimposing a dc current to the device through bias tees at room temperature. The dc current is slowly increased while monitoring the resistance of the transmission line. The resulting critical current is larger than $1 \, \mathrm{mA}$.

Further preliminary measurements have been conducted at millikelvin temperature. For this, a full yet preliminary set-up comprising cold attenuators, bias-tees and isolators hosted in a dilution refrigerator with a base temperature of $20 \, \mathrm{mK}$ has been employed.
The scaling current $I_*$ has been estimated through phase measurements as function of the dc current.
In this setting, a dc current $I_{dc}$ is superimposed to the weak VNA rf signal.
The output phase shift is then measured as function of $I_{dc}$, as shown in figure \ref{SUBFIG:DWD2_phase_shift}. 
The signal frequency is set slightly above the bandgap. Since the bandgap corresponds to stationary waves, with half wavelength equal to the length a \textit{super-cell}, the total phase shift along the device can be estimated as $\theta_0 \approx N_{sc}\, \pi$ where $N_{sc}=523$ is the total number of \textit{super-cells}.

The theoretical relative phase shift is given by:

\begin{equation}\label{eq:phaseshift}
    \frac{\Delta \theta}{\theta_0} = \frac{1}{2}\left( \frac{I_{dc}}{I_*}\right)^2\quad\cite{Gao2014}
\end{equation}

\noindent and from a linear fit of $\Delta \theta/\theta_0$ as a function of $I_{dc}^2$ it is possible to obtain the scaling current $I_*=(5.3 \pm 0.1)\,$mA. This value is in reasonable agreement with the theoretical value that is predicted to be $3/2\,\sqrt{3}\,I_c$ \cite{Zmuidzinas2012}, where $I_c = (1.5 \pm 0.2)$ mA is the critical current.

A very preliminary functional test has been performed on the device as a parametric amplifier by pumping it with a pump frequency $f_p$ slightly above the bandgap. 
The KI-TWPA device performance has been tested for signal frequencies close to  $f_p/2$, showing very promising results and pointing towards an achievable gain up to 11$\,$dB.
A complete cryogenic characterisation with a dedicated set-up is planned after the fabrication of the KI-TWPA device with the final length of about 33$\,$cm.

This preliminary characterisation results clearly demonstrate that the developed NbTiN film is suitable for the development of KI-TWPA devices with high gain and potentially for other kind of devices based on high kinetic superconductors.

\section{Conclusion and outlook}

In this study, we have presented the optimisation of the rf sputter-deposition process of NbTiN films using a Nb$_{80\%}$Ti$_{20\%}$ target, with the goal of achieving precise control over film characteristics, resulting in high kinetic inductance and high linearity while maintaining a high transition temperature. Through the exploration of the parameter landscape related to the different sputtering conditions, such as pressure, power, and nitrogen flow, and the use of film thickness as a fine-tuning parameter, we were able to adjust the properties of the final NbTiN films, resulting in a kinetic inductance of 8.5$\,$pH/sq, a value that is highly suitable for our applications. 
A first KI-TWPA prototype fabricated from the optimised film underwent a preliminary characterisation, showing the expected response, the expected non-linearity and promising gain performance.

The results of this study provide a solid foundation for further developments and optimisations of Kinetic Inductance Travelling Wave Parametric Amplifiers (KI-TWPAs) within the DARTWARS project and beyond. The final KI-TWPA device with a length of about 33$\,$cm will be designed, fabricated and fully characterised with a dedicated cryogenic set-up tailored to the DARTWARS experiment. Additionally, the research here presented can be further extended to explore new materials and fabrication processes in order to maximise the amplifier performances. With the increasing interest in quantum technologies, the development of high-performance amplifiers is of paramount importance and the results of this study are expected to contribute to the advancement of this field.

\section*{Data availability statement}
The data that support the findings of this study are available upon reasonable request from the authors.

\section*{Acknowledgement}
This work is supported by DARTWARS, a project funded by the European Union’s H2020-MSCA Grant Agreement No. 101027746 and by the Italian Institute of Nuclear Physics (INFN) within the Technological and Interdisciplinary Research Commission (CSN5). The work is also supported by the Italian National Quantum Science and Technology Institute through the PNRR MUR Project under Grant PE0000023-NQSTI. We acknowledge the support of the FBK cleanroom team for the fabrication. We also acknowledge useful discussions with Jiansong Gao, Michael Vissers, and Jordan Wheeler.

\section*{References}

\bibliographystyle{iopart-num}
\bibliography{biblio}

\end{document}